# Exchange operation of Majorana zero modes in topological insulator-based Josephson trijunctions

Yunxiao Zhang[1,2], Zhaozheng Lyu[1,2,4,†], Xiang Wang[1,2], Yukun Shi[1,2], Duolin Wang[1,2], Xiaozhou Yang[1,2], Enna Zhuo[1,2], Bing Li[1,2], Yuyang Huang[1,2], Zenan Shi[1,2], Anqi Wang[1,2], Heng Zhang[3], Fucong Fei[3], Xiaohui Song[1,4], Peiling Li[1], Bingbing Tong[1], Ziwei Dou[1], Jie Shen[1], Guangtong Liu[1,4], Fanming Qu[1,2,4], Fengqi Song[3,†] and Li Lu[1,2,4,†]

[1] *Beijing National Laboratory for Condensed Matter Physics, Institute of Physics, Chinese Academy of Sciences, Beijing 100190, China*

[2] *School of Physical Sciences, University of Chinese Academy of Sciences, Beijing 100049, China*

[3] *College of Physics, Nanjing University, Nanjing 210008, China*

[4] *Hefei National Laboratory, Hefei 230088, China*

† Corresponding authors: lyuzhzh@iphy.ac.cn, songfengqi@nju.edu.cn, lilu@iphy.ac.cn

**Abstract**

Majorana zero modes are anyons obeying non-Abelian exchange statistics distinct from fermions or bosons. While significant progresses have been achieved in the past two decades in searching for these exotic excitations in solid-state systems, their non-Abelian nature remains unverified, as definitive proof requires braiding operations. Here, we report preliminarily experimental advances in creating, manipulating, and exchanging the presumed Majorana zero modes in an envelope-shaped Josephson device composed of multiple trijunctions on a topological insulator surface. We observed the signatures of in-gap states migration consistent with the expectations of the Fu-Kane model, supporting the realization of an exchange operation. This work would establish a critical pathway toward ultimately braiding Majorana zero modes in the Fu-Kane scheme of topological quantum computation.

In 1982, Richard Feynman pointed out that a quantum machine could better simulate the real physical world whose underlying mechanism is governed by quantum mechanics [1]. The inherent superposition and entanglement of quantum states enable parallel computation, offering superior efficiency for solving certain NP-hard problems compared to classical computers. To achieve quantum computation, diverse physical systems are being developed to host quantum bits (qubits) with sufficiently long coherence times. Among these, platforms supporting non-Abelian anyons [2-4]—such as Majorana zero modes (MZMs)—hold promise for hosting topologically protected qubits due to inherent fault tolerance. Even within the field of topological quantum computation there are a number of key schemes, including the ones based on semiconductor nanowires with strong spin-orbit coupling [5-10], atomic chain [11], artificial Kitaev chains [12-14], vortex cores of topological insulator (TI)-based [15-17] and iron-based superconductors [18-20], Josephson junctions (JJs) constructed on TIs [15, 21-25], etc. For the last scheme, Fu and Kane [15] predicted in 2008 that a Josephson trijunction on a 3D TI surface hosts an MZM at its center when the minigap $\delta_j = \Delta \cos(\varphi_j/2)$ becomes negative in an odd number of single junctions. Here, $\varphi_j$ ($j = 1, 2, 3$) is the superconducting phase difference across the $j^{\text{th}}$ single junction, and $\Delta$ is the proximity-induced gap. Subsequent proposals suggest braiding MZMs by tuning superconducting phases across multiple trijunctions, enabling universal quantum computation via surface code protocols in trijunction arrays [26].

Previously, the MZM phase diagram of trijunction predicted by Fu and Kane has been experimentally verified [22], and the coupling between MZMs in two neighboring trijunctions has also been observed [24]. In this work, aiming to ultimately braid the MZMs, we further explored the exchange operation of two MZMs along predefined paths in a Josephson device containing four trijunctions.

Figure 1(a) shows the scanning electron microscopic image of the device. It contains four Ti/Al superconducting pads fabricated on a ~30 nm thick exfoliated flake of Sn-Bi$_{1.1}$Sb$_{0.9}$Te$_2$S (SBSTS), forming an envelope-shaped device with six single junctions and four trijunctions. The single junctions have a same length of ~170 nm, but with various widths and thus various critical supercurrents ranging up to ~300 nA at the base temperature of ~40 mK in this experiment. The ring-shaped outer pad helps to screen unwanted coupling and noise from the environment during MZMs manipulation.

To allow flux control of the phases of the superconducting pads, thus the phase differences of the junctions, three superconducting loops were attached to connect between the inner pads and the outer ring (Fig. 1(c)), with $S_1 \approx 5~\mu m^2$ and area ratios of $S_1:S_2:S_3 = 1:(-4):4$ (negative means reversed direction). The phase difference across a single junction is solely determined by the magnetic flux in the loop that connects to that junction. If let the outer superconducting ring being the phase reference, the phases of the other three pads driven by a magnetic field $B$ are $\phi, -4\phi, 4\phi$, respectively, as labeled in the first panel of Fig. 1(e), where $\phi = 2\pi BS_1/\phi_0$ and $\phi_0$ is the flux quantum. In the same figure, the phase differences $\varphi$ across each single junction is integer multiples of $\phi$ represented by the numbers of black squares in corresponding junction, with which the minigap of the junction is determined.

In the following, we use $\phi$ to represent the magnetic field. Obviously, no MZM exists in the interval $\phi \in (0, \pi/8)$. When $\phi > \pi/8$, the phase difference across the central single junction satisfies $\varphi > \pi$, so that the corresponding minigap $\Delta \cos(\varphi/2)$ becomes negative (with junction area marked in light grey in Fig. 1(e)), and a pair of MZMs ($\gamma_1, \gamma_2$) nucleates at the two ends of the junction. This state keeps for $\phi \in (\pi/8, \pi/5)$. Beyond $\phi = \pi/5$, the minigap of the upper left single junction becomes negative, so that $\gamma_1$ migrates to the center of the upper left trijunction. With the further increase of magnetic field, the device undergoes sequential topological transitions when crossing from the interval $(\pi/5, \pi/4)$ to $(\pi/4, \pi/3)$ to $(\pi/3, 3\pi/8)$ to $(3\pi/8, 3\pi/5)$ one by one, accompanied with the migrations of $\gamma_1$ and $\gamma_2$ along predefined paths depicted in Fig. 1(e). Beyond $\phi = 3\pi/8$, the minigap of the central junction becomes positive again, so that the MZMs pair is annihilated/fused.

To trace the positions of MZMs in the device, we further fabricated a number of Al/Au normal-metal probe electrodes (colored in yellow in Fig. 1(a)). The tips of these electrodes touched with $Ar_2/O_2$ plasma-cleaned SBSTS surfaces at the centers of the trijunctions, such that local contact conductance measurement, which probes mostly the surface states, could be performed. The 3 nm-thick Al layer reacts with SBSTS to form a tunneling barrier [27]. The measurements were carried out with standard low frequency lock-in technique in a dilution refrigerator.

Among the six normal-metal probe electrodes, only three of them at positions $P_1$, $P_2$ and $P_3$ were functioning. Shown in Fig. 2 are the data of contact conductance $dI/dV$ measured at $P_1$ and $P_2$ as functions of dc bias voltage $V_{dc}$ and $\phi$. The data measured at $P_3$ are presented in the Supplemental Material. The measurement configuration is illustrated in Fig. 1(b). During the measurements the ac voltage modulation was $V_{ac} = 20\ \mu V$. From the 2D maps and the line cuts one can see that the contact conductance exhibits a soft gap-like structure which oscillates with the variation of magnetic field.

To understand the physics behind the experimental data, we calculated the Andreev bound states (ABSs) based on a lattice model for the trijunctions. According to the Fu-Kane theory [15] and the Su-Schrieffer-Heeger model [28], we discretized each trijunction into a lattice with $3N$ sites (Fig. 3(a)) whose Hamiltonian is [29]: $H_{eff} = i\hbar v_M(\gamma_L \partial_y \gamma_L - \gamma_R \partial_y \gamma_R) + i\Delta \cos(\varphi_j/2)\gamma_L \gamma_R$, where $t = \hbar v_M/(2a_0)$ is the nearest-neighbor hopping term, $v_M$ is the Majorana velocity, and $a_0$ the distance between neighboring lattice sites. We choose the total number of sites to be $3N = 600$. The results of simulation do not change much if a different total number of sites is chosen. The diagonalization of the lattice Hamiltonian yields $H_{tri} = \sum_n \varepsilon_n \psi_n^+ \psi_n$, where $\varepsilon_n$ ($n = 1$ to 600) is the eigenenergy of the $n^{th}$ ABS band. The detailed calculation was explained in the Supplemental Materials of Ref. [22].

In Figs. 3(d, g) we present the calculated ABS bands as a function of magnetic field for the trijunctions centered at $P_1$ and $P_2$, respectively. The bands of the in-gap states are displayed in red ($n = 1, 2$), and the bands of the continuum ABS states are displayed in blue ($n = 3$ to 50) and black ($n = 51$ to 600). Figures 3(b, c) show that the in-gap states are mainly contributed and mostly localized at the trijunction centers, whereas the continuum states are mainly contributed by the single junctions, forming a series of standing waves in the single junctions but with suppressed local density of states (LDOS) at the trijunction center. And Figs. 3(e, f, h, i) show that, at the trijunction centers, the $\phi$ dependence of the LDOS of the in-gap states is distinctly different from that of the continuum states—when one is maximized, the other is minimized. This feature is used to help allocating the contribution of the in-gap states to the measured contact conductance.

According to the Fu-Kane model, the in-gap states in the $0^{th}$ white lobes centered at $\phi = 0$ in Figs. 3(d, g) are topologically trivial, and the ones in the $1^{st}$ and the $2^{nd}$ lobes in Fig. 3(d), as well as the ones in the $1^{st}$ and the $3^{rd}$ lobes in Fig. 3(g), though their energies are not exactly zero, are topologically nontrivial and MZM-related. We call them topological nontrivial in-gap states and mark their central positions in Figs. 3 and 4 with dashed vertical lines. We note that the finite energy of these nontrivial modes arises from the finite size of the trijunction, as detailed in the Supplemental Materials of Ref. [22]. Furthermore, the six red lattice sites in the central circle (Fig. 3(a)) deviate slightly from the exact center of the trijunction, which may also pick up spatial energy oscillation of the modes—similar to that observed near the ends of a nanowire [30, 31].

To further simulate the data of contact conductance based on the lattice model, we need to analyze the electron transport process between the trijunction and the probe electrodes. The Hamiltonian of the electrons in the probe electrode is [32]: $H_{\text{probe}} = \sum_k \varepsilon_k C_k^+ C_k$, and the tunneling Hamiltonian is $H_T = \sum_{m \in \text{circle}} \tau C_k^+ a_m + h.c.$, where the summation is over the six red sites within the dashed circle directly beneath the probe electrode, and the coupling strength $\tau$ is set to constant for these six lattice sites. Through unitary basis transformation $a_m = \sum_n \langle m|\psi_n\rangle \psi_n$, the tunneling Hamiltonian becomes $H_T = \sum_n \tau^* C_k^+ \psi_n + h.c.$, where $\tau^* = \sum_{m \in \text{circle}} \langle m|\psi_n\rangle$ is the tunneling matrix element. The magnitude of $\tau^*$ is directly related to the local wavefunction density of the eigenstate $\psi_n$ beneath the probe electrode.

According to the expression of contact conductance of the probe electrodes in the tunneling limit [33], we have: $dI/dV \propto |\tau^*|^2 \rho_{\text{probe}}(E_F) \rho_{\text{tri}}(E_F + eV)$, where $\rho_{\text{tri}}(\epsilon) = -\frac{1}{\pi} \text{Im} G^r(\epsilon) = \sum_n \frac{1}{2\pi} \frac{\Gamma}{[\epsilon-\varepsilon_n]^2+[\Gamma/2]^2}$ is the LDOS of the trijunction dominated by the in-gap states, $G^r$ is the retarded Green's function, $\Gamma \propto |\tau^*|^2 \rho_{\text{probe}}(E_F)$ is the broadening of the energy levels [34] as illustrated in Fig. 1(d), $\rho_{\text{probe}}(E_F)$ is the LDOS of the probe electrode near the Fermi level $E_F$, $\epsilon = eV$, and $\varepsilon_n$ is the energy of the $n^{th}$ ABS band given by the lattice model. Our measured contact conductance is in the tunneling regime, because the conductance inside the minigap is suppressed, we therefore follow the above formalism to propose a simplified fitting formula containing two $\phi$-dependent fitting parameters $\alpha$ and $\Gamma$, to fit the measured contact conductance in a given magnetic field:

$$dI/dV = \alpha \sum_n \frac{\Gamma}{[\epsilon - \varepsilon_n]^2 + [\Gamma/2]^2} \qquad (1)$$

where the level broadening $\Gamma$ is proportional to $|\tau^*|^2$, and $\alpha$ represents the probe electrode's averaged tunneling strength to the trijunction which, according to the above analysis, is also proportional to $|\tau^*|^2$ [35]. Although $\alpha$ and $\Gamma$ exhibit a similar behavior to certain extent, they cannot be merged into a single fitting parameter—while we assume identical broadening ($\Gamma$) for all ABS bands due to rapid thermal equilibrium, the spatial weight of different bands at the trijunction center, and consequently of their tunneling matrix elements, varies differently with $\phi$.

By simulating the data taken at $P_1$ along the violet-colored vertical line cut at $\phi = -0.25 \times 2\pi$ in Fig. 2(a) (the data are shown as violet-colored dots in Fig. 2(b)), we obtained the lower black curve in Fig. 2(b), with fitting parameters $\Delta_1 \approx 167\ \mu eV$, $t_1 \approx 133\ \mu eV$, and a set of $(\alpha, \Gamma)$ values at that $\phi$. Then, by fixing $\Delta_1$ and $t_1$ as two global fitting parameters, we were able to simulate the whole 2D map of the measured contact conductance in Fig. 2(a) with fitting parameters $(\alpha, \Gamma)$ at each $\phi$. The result of simulation is shown in Fig. 4(a), and the $\phi$ dependences of $\alpha$ and $\Gamma$ are shown in Figs. 4(b, c). Similar simulations can be carried out for the data taken at $P_2$, with $\Delta_2 \approx 71\ \mu eV$, $t_2 \approx 151\ \mu eV$ and sets of $\phi$-dependent $(\alpha, \Gamma)$ shown in Figs. 4(e, f). It can be seen that the simulated 2D maps in Figs. 4(a, d) and the line cuts in Figs. 2(b, c, e, f) well reproduce the peculiar patterns of the measured contact conductance, including the wiggles in the horizontal line cuts. The obtained large level broadening $\Gamma$ also self-consistently explains why the calculated topologically nontrivial structures centered at the dashed vertical lines in Figs. 3(d, g) are missing in the measured data. The results indicate that the lattice model and the level broadening picture used in the analysis are validate for our device.

Let us further discuss the origin of level broadening and its consequence. If we were in the $\Gamma \to 0$ limit, $\rho_{tri}(\epsilon)$ would be reduced to $\rho_{tri}(\epsilon) = \sum_n \delta(\epsilon - \varepsilon_n)$, so that the spectral weight would be concentrated strictly at the eigenenergies $\varepsilon_n$, enabling clear resolution of discrete bands and the zero-bias conductance peak (ZBCP) in $dI/dV$ if existed. However, given the fact that the contact conductance in this experiment was about ten times larger than the quantum unit $e^2/h$, we were in the $\Gamma \gg \varepsilon_{n+1} - \varepsilon_n$ regime, so that the in-gap states were significantly broadened and immersed in the continuum states. This is because $\Gamma \propto |\tau^*|^2 \rho_{\text{probe}}(E_F)$ and $\tau^* = \sum_{m \in \text{circle}} \langle m | \phi_n \rangle \tau$, $\Gamma$ is directly linked to the LDOS beneath the probe electrode. The coupling between the electrons in the probe electrode and those in the trijunction center reduces the lifetime of the electrons in the latter, broadening their levels to a Lorentzian-shaped distribution [36]. On the other side, while the broadening $\Gamma$ increases with the coupling strength [37, 38], the central energy of the levels remains unshifted. This is true even at large $\Gamma$ (such as $\Gamma/\Delta > 5$), as revealed in simulating the MZMs at vortices core in STM experiments [39]. The study of ZBCP broadening in InAs-related experiment also yields similar conclusion [40].

Although the low energy resolution prevented us from resolving ZBCPs as signatures of MZMs, the high spatial resolution of the local probe technique, with a tip size of ~50 nm, together with the expected particular $\phi$ dependence of the in-gap states (Figs. 3(e, h)), would enable us to resolve tiny LDOS change when an in-gap state migrates into or away from the area beneath the probe electrodes during the variation of $\phi$. In the following we will show that, the tiny LDOS change indeed influences the amplitude of the conductance and the broadening of the ABS levels, causing corresponding changes in the fitting parameters $\alpha$ and $\Gamma$, thereby allowing us to tracking the migration of the in-gap states in the devices as $\phi$ varies.

The black curves in Figs. 4(b, c, e, f) are the $\phi$ dependences of $\alpha$ and $\Gamma$ obtained in simulating the $dI/dV$ data in Figs. 2(a, d) measured at positions $P_1$ and $P_2$, respectively. Also shown as red curves in Figs. 4(b, c, e, f) are the $\phi$ dependences of the LDOS of the in-gap states predicted by the lattice model. It can be seen that the $\phi$ dependences of $\alpha$ and $\Gamma$ are positively correlated to the LDOS of the in-gap states beneath the probe electrodes—every time when the wavefunctions of the in-gap states concentrate at the trijunction center, both $\alpha$ and $\Gamma$ exhibit a peak. It is straightforward to understand that the coefficient $\alpha$ is directly proportional to the LDOS of the in-gap states. As for the coincidence between the $\Gamma$ peaks and the LDOS peaks of the in-gap states, the results support the scenario that the coupling between the probe electrode and the ABS states in the trijunction firstly broadens the in-gap states whose majority weight is directly beneath the probe electrodes at the trijunction center, then the broadening spreads to the continuum states whose majority weight is in the single junctions.

In Figs. 4(b, c, e, f), the peaks of $\alpha$ and $\Gamma$ are slight misaligned from those of the LDOS of the in-gap states. The misalignment might arise from two modifications: (i) the modification of effective magnetic flux in the superconducting loops $(S_1, S_2, S_3)$ caused by inductive screening; (ii) the modification of effective area of each superconducting loop caused by the Meissner effect of the superconducting pads. After taking into account these modifications, we show in the Supplemental Material that the peaks in $\alpha$ and $\Gamma$ are better aligned with the peaks of the LDOS of the in-gap states.

Finally, let us discuss the $\phi$-depenent sequential topological transitions of the device as illustrated in Fig. 1(e). The different intervals of $\phi$ in Fig. 1(e) are marked in Figs. 4(b, c, e, f) with the same colors. According to the Fu-Kane model, when entering from interval ①(pink) to ②(yellow), a pair of $\gamma_1$ and $\gamma_2$ is nucleated. Within the intervals ②(yellow) and ③(green), there will be $\gamma_2$ at position $P_1$, but no MZM at position $P_2$. Indeed, there is a broad peak in the ②(yellow) and ③(green) intervals in both the fitting parameters ($\alpha$, $\Gamma$) and in the calculated LDOS of the in-gap states for position $P_1$ (Figs. 4(b, c)), but no peak in the same intervals for position $P_2$ (Figs. 4(e, f)). With further increasing $\phi$ to interval ④(blue) and then to interval ⑤(violet), $\gamma_1$ migrates to $P_1$, giving rise to the broad peak in the ④(blue) and ⑤(violet) intervals in Figs. 4(b, c). Meanwhile, $\gamma_2$ migrates to $P_2$ in interval ④(blue), then further migrates away from $P_2$ to the place where $\gamma_1$ was in interval ⑤(violet), causing the broad peak near interval ④(blue) in Figs. 4(e, f). By doing so, an exchange operation between $\gamma_1$ and $\gamma_2$ is completed.

To summarize, based on the Fu-Kane model, we designed and fabricated an envelope-shaped Josephson device composed of multiple single junctions and trijunctions. By varying the magnetic field, we demonstrated the creation, manipulation, and exchange operation of the in-gap states at the trijunction centers. Although our experiment lacked the energy resolution to resolve ZBCP-related features of the nontrivial in-gap states, due to level broadening caused by the probe electrodes, with the high spatial resolution of the local-probe conductance spectroscopy and combined with the lattice model, we were able to capture their signatures of migration along predefined paths through the $\phi$ dependences of fitting parameters. It has to be pointed out that the tunneling measurement which aims for detecting and tracking the MZMs not only broadens the levels of the states, but also directly poisons the parity of the system by injecting quasiparticles [41, 42], therefore should eventually be abandoned. A superior strategy in the future would involve inductive detection of the parity-dependent supercurrent via high-bandwidth single-shot measurements. Such an approach would significantly reduce direct coupling to the MZMs and allow for fast, coherent readout within the quasiparticle poisoning timescale. Our previous experiments on Sb-$Bi_2Te_3$ nanowire-based Transmons revealed a quasiparticle poisoning time of $\tau_{qp} \sim 0.5\ \mu s$ without employing special protection [43]—a timescale which is feasible for performing coherently braiding by using mature radio-frequency techniques.

**Acknowledgements** The authors wish to thank Yuval Oreg for providing helpful suggestions. This work was supported by the Innovation Program for Quantum Science and Technology through Grant No. 2021ZD0302600; by NSFC through Grant Nos. 92065203, 92365302, 11527806, 12074417, 11874406, 11774405, E2J1141, 92161201, 12374043 and 12474272; by the Strategic Priority Research Program B of the Chinese Academy of Sciences through Grants Nos. XDB33010300, XDB28000000, and XDB07010100; by the National Basic Research Program of China through MOST Grant Nos. 2016YFA0300601, 2017YFA0304700, 2015CB921402 and 2022YFA1402404; by Beijing Natural Science Foundation through Grant No. JQ23022; by Beijing Nova Program through Grant No. Z211100002121144; and by Synergetic Extreme Condition User Facility (SECUF).

**Author contributions** Y.Z., Z.L. and L.L. conceived the experiment. Y.Z. fabricated the devices and performed the measurements. Y.Z. and Z.L. analyzed the data and did the numerical simulations. H.Z., F.F. and F.S. provided the materials, Y.Z. and L.L. wrote the manuscript. All the authors participated in the discussion.

**Supplemental Material** is available online.

**Competing interests** The authors declare no competing financial interests.

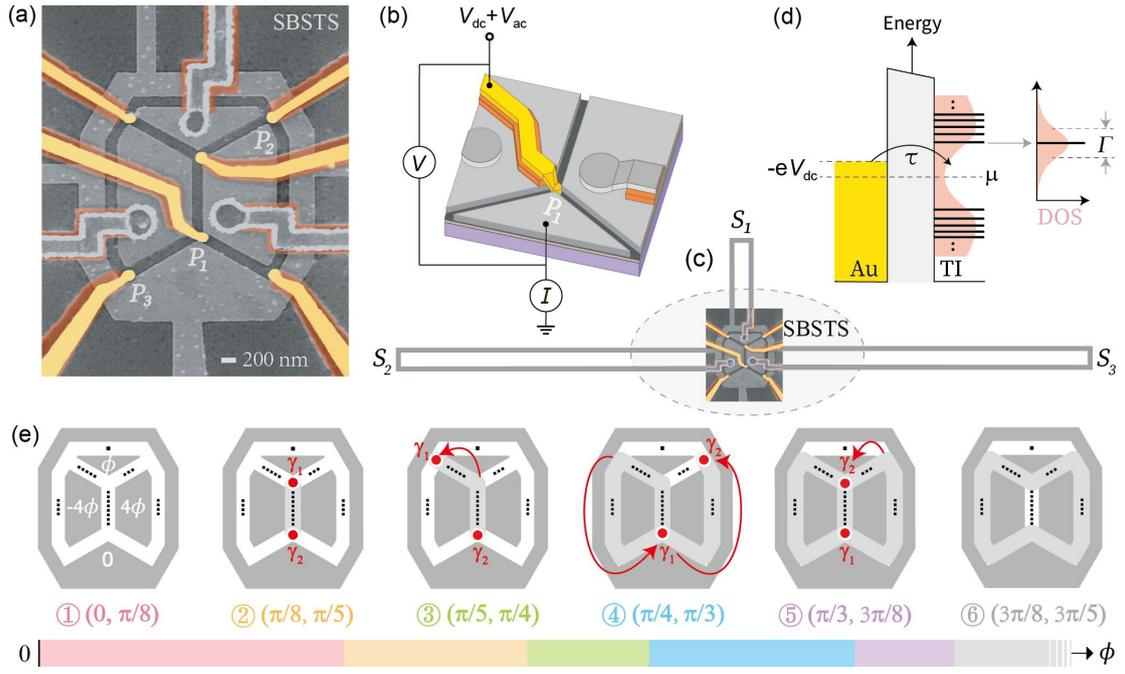

FIG. 1. (a) False-color SEM image of the envelope-shaped Josephson device with six single junctions and four trijunctions, composed of exfoliated SBSTS flake (black), Ti/Al superconducting pads (grey), Al/Au normal-metal probe electrodes (yellow), and $Al_2O_3$ insulating layer (brown). (b) Schematic of the device structures near $P_1$, together with the illustration of contact conductance measurement. (c) Schematic of the device with three external superconducting loops. (d) Illustration of the broadening of ABS levels due to coupling with the probe electrodes, which smears the discrete levels to a continuum of local density of states (LDOS, orange). (e) Sequential topological transitions of the device in a magnetic field. The number of black squares in the single junctions represent the amount of phase difference across the junctions generated by the magnetic field. With the increase of magnetic field, the minigap of the single junctions becomes negative (light grey) one by one when crossing the boundaries between the intervals represented by different colors, which creates/braids/annihilates a MZMs pair.

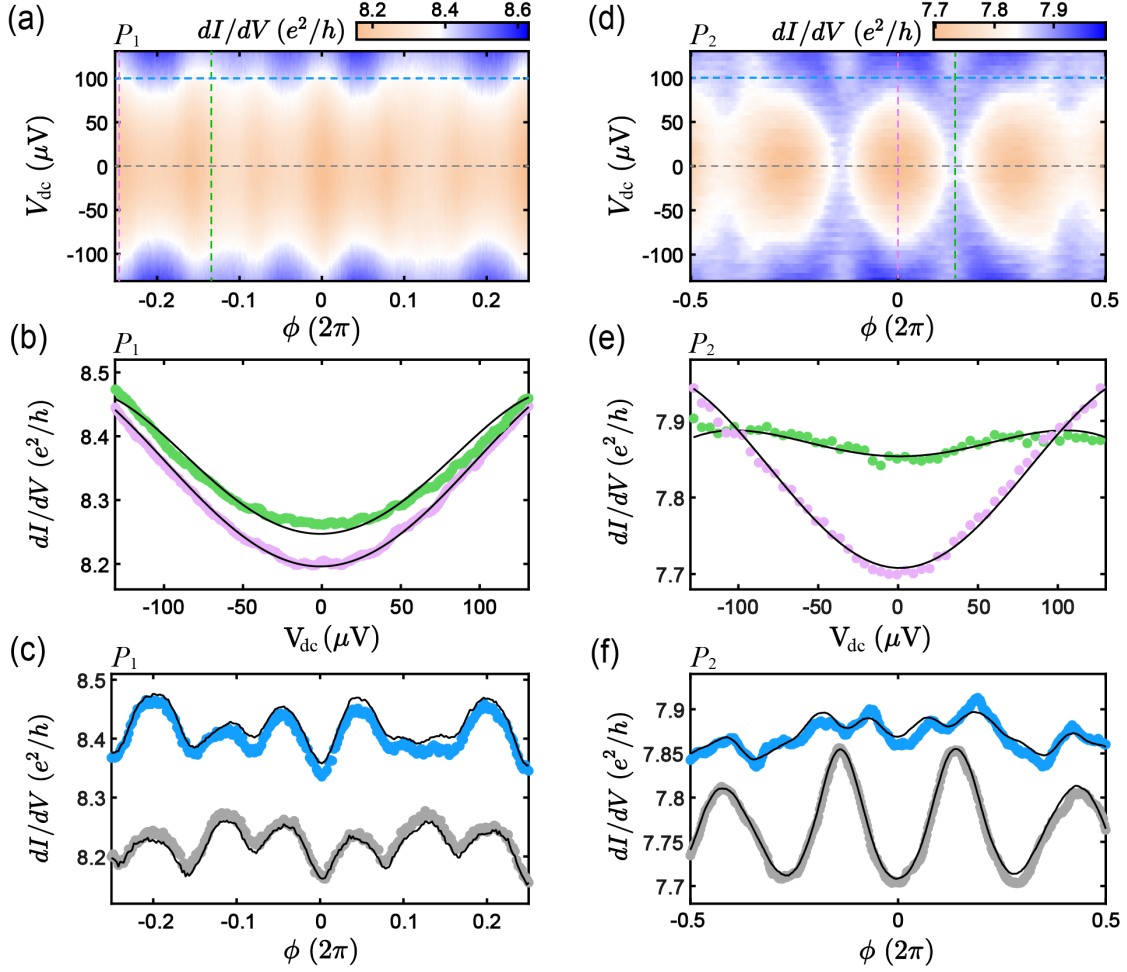

FIG. 2. (a) 2D map of contact conductance measured at position $P_1$ as a function of dc bias voltage and $\phi$. (b) Vertical line cuts in (a), at $\phi = -0.25 \times 2\pi$ in the topologically trivial interval (violet dots) and at $\phi = -0.125 \times 2\pi$ in the topologically nontrivial interval (green dots). (c) Horizontal line cut in (a) at $V_{dc} = 0$ (grey dots) and at $V_{dc} = 100\,\mu V$ (blue dots). (d) 2D map of contact conductance measured at $P_2$ as a function of dc bias voltage and $\phi$. (e) Vertical line cuts in (d), at $\phi = 0$ in the topologically trivial interval (violet dots) and at $\phi = 0.146 \times 2\pi$ in the topologically nontrivial interval (green dote). (f) Horizontal line cut in (d) at $V_{dc} = 0$ (grey dots) and at $V_{dc} = 100\,\mu V$ (blue dots). The black curves in (b, c, e, f) are the results of numerical simulation with the model and parameters presented in Figs. 3 and 4.

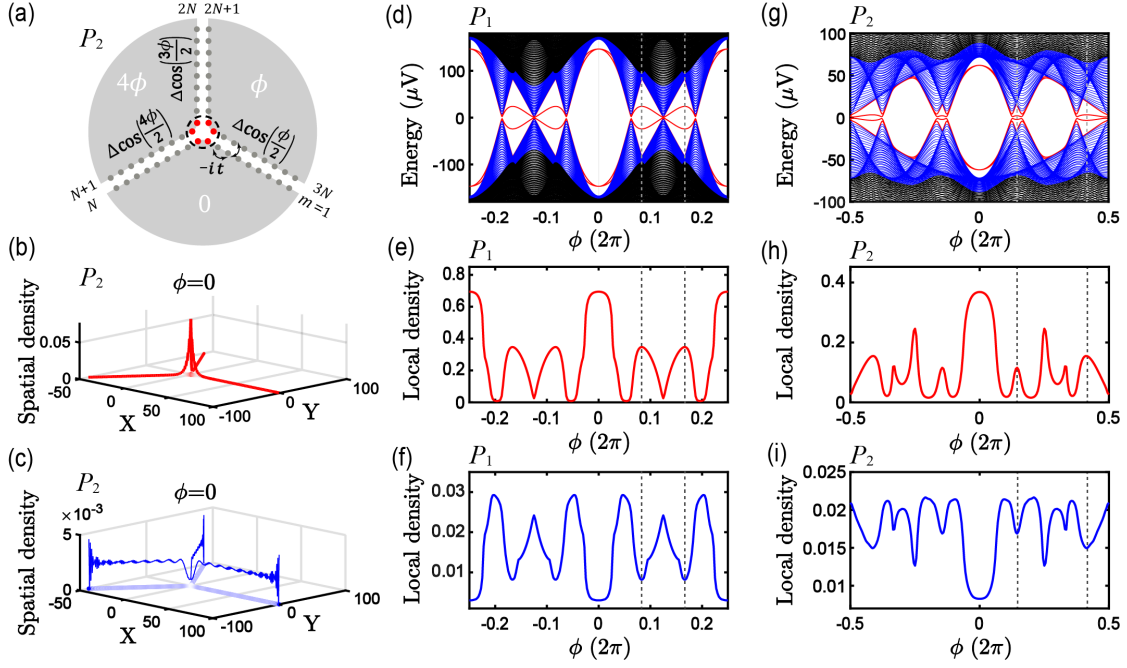

FIG. 3. (a) The lattice model for the trijunction at position $P_2$, with specific phase differences in a magnetic field determined by the geometry, and with the minigap $\delta = \Delta\cos(\varphi/2)$ and the nearest-neighbor hopping amplitude $t$ as two global fitting parameters. (b, c) Spatial density distributions of the trijunction's in-gap states (red: $n = 1, 2$) and the single junctions' continuum states (blue: $n = 3$ to $50$), respectively, at $P_2$ and when $\phi = 0$. (d) The magnetic field dependence of the calculated ABS bands for the trijunction at $P_1$. Illustrated in red are the in-gap states (band indies $n = 1, 2$) of the trijunction, and in blue ($n = 3$ to $50$) and black ($n = 51$ to $600$) are the continuum states of the single junctions. The in-gap states in the $0^{th}$ white lobe centered at $\phi = 0$ are topologically trivial, and in the $1^{st}$ and the $2^{nd}$ white lobes (marked by dashed vertical lines) are topologically nontrivial. (e, f) The magnetic field dependence of the local density at the trijunction center for the in-gap states (red: $n = 1, 2$) and the continuum states (blue: $n = 3$ to $50$ as the representatives). Note that counting the contributions of the bands from $n = 51$ to $600$ make no big difference. (g-i) Similar numerical simulations for the trijunction located at $P_2$. The in-gap states in the $0^{th}$ and the $2^{nd}$ white lobes are trivial, and in the $1^{st}$ and $3^{rd}$ lobes are nontrivial.

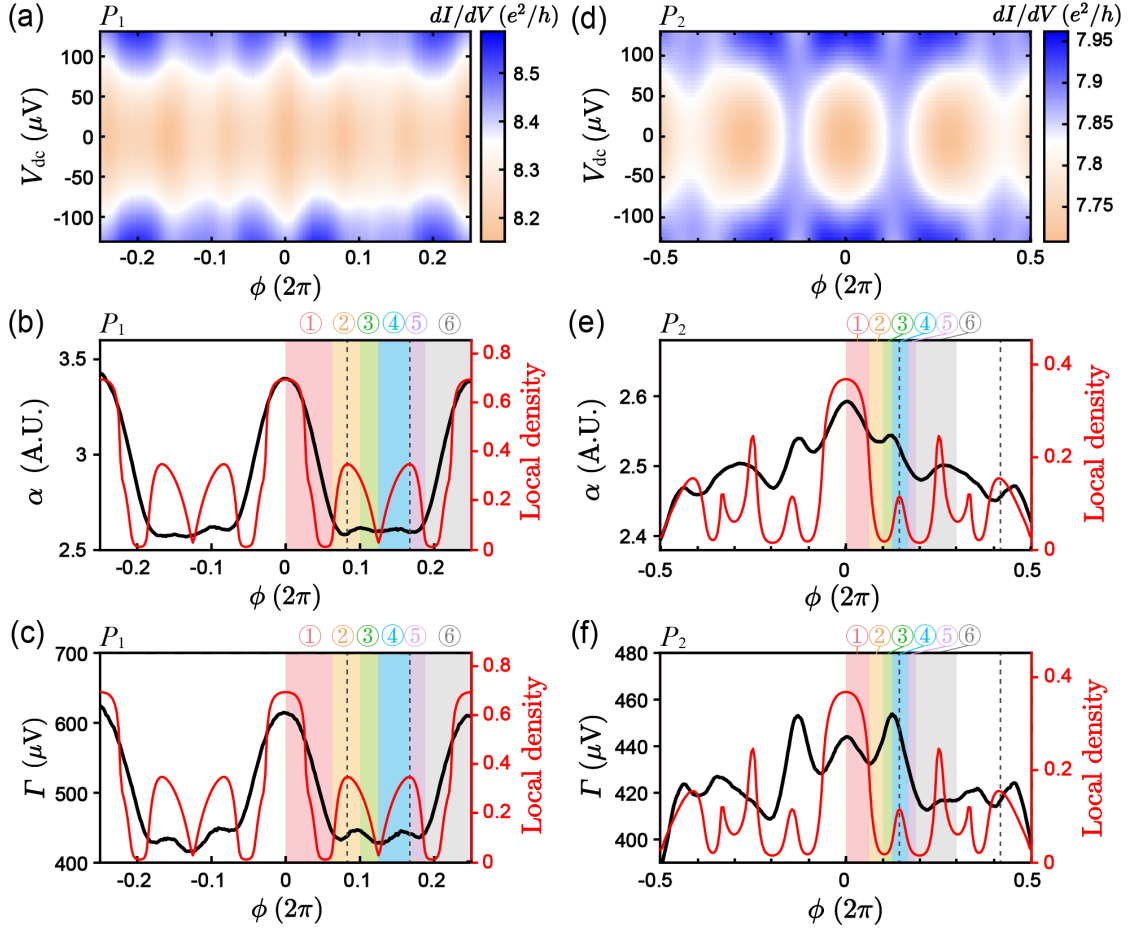

FIG. 4. (a) The simulated $\phi$ dependence of 2D map of contact conductance at position $P_1$. (b) The $\phi$ dependences of fitting parameter α (black) and the calculated local density of in-gap states (red) at $P_1$. The sequential intervals of $\phi$ illustrated in Fig. 1(e) are marked with the same colors here. α maximizes at $\phi = 0.084 \times 2\pi$ and $0.167 \times 2\pi$, close to the positions of local density peaks of the topologically nontrivial in-gap states indicated by the dashed vertical lines. (c) The $\phi$ dependences of fitting parameter $\Gamma$ (black) and the calculated local density of the in-gap states (red) at $P_1$, also demonstrating a positive correlation between them. (d-f) Similar results for the trijunction located at $P_2$: α and $\Gamma$ maximize at $\phi = 0.146 \times 2\pi$ and $0.416 \times 2\pi$, close to the positions of the local density peaks of the topologically nontrivial in-gap states, demonstrating a positive correlation again. Better alignment between the expected and the fitted peak positions can be achieved after taking into account the modifications of magnetic flux due to Meissner screening of the superconducting pads and rings [Supplemental Material].

# Supplemental Material for "Exchange operation of Majorana zero modes in topological insulator-based Josephson trijunctions"

Yunxiao Zhang[1,2], Zhaozheng Lyu[1,2,4,†], Xiang Wang[1,2], Yukun Shi[1,2], Duolin Wang[1,2], Xiaozhou Yang[1,2], Enna Zhuo[1,2], Bing Li[1,2], Yuyang Huang[1,2], Zenan Shi[1,2], Anqi Wang[1,2], Heng Zhang[3], Fucong Fei[3], Xiaohui Song[1,4], Peiling Li[1], Bingbing Tong[1], Ziwei Dou[1], Jie Shen[1], Guangtong Liu[1,4], Fanming Qu[1,2,4], Fengqi Song[3,†] and Li Lu[1,2,4,†]

[1] Beijing National Laboratory for Condensed Matter Physics, Institute of Physics, Chinese Academy of Sciences, Beijing 100190, China

[2] School of Physical Sciences, University of Chinese Academy of Sciences, Beijing 100049, China

[3] College of Physics, Nanjing University, Nanjing 210008, China

[4] Hefei National Laboratory, Hefei 230088, China

† Corresponding authors: lyuzhzh@iphy.ac.cn, songfengqi@nju.edu.cn, lilu@iphy.ac.cn

## Contents

1. Corrections to the peak positions in Fig. 4 (main manuscript) incorporating the Meissner effect and the self-inductance of the superconducting loops
2. The contact conductance measured by the probe electrode at position $P_3$ of the envelope-shaped device

# 1. Corrections to the peak positions in Fig. 4 (main manuscript) incorporating the Meissner effect and the self-inductance of the superconducting loops

In the main manuscript, the slight offsets between the peak positions of the $\phi$-dependent fitting parameters ($\alpha, \Gamma$) and the local density of in-gap states can be explained by taking into account the Meissner screening of the superconducting pads and inductance effect of the superconducting loops. There are two sources that will lead to an effective magnetic flux in the superconducting loops different from the applied one.

Firstly, we consider the redistribution of the magnetic flux caused by the Meissner effect of the superconducting pads. Since flux lines cannot penetrate superconductors below the lower critical magnetic field, the magnetic flux that would originally pass through the superconducting pads is expelled and enters the loops. The area corresponding to these expelled regions should be included in the area of the loops, which leads to a new effective area $S_{\text{eff}}$. Here, we use the method from the Supplemental Materials of [21, 22] for a simple estimation: considering the magnetic flux originally heading to the area defined by the ring-shape loop of the device (the board of the envelope). The repelling results in the following changes in the loop areas: $S_{1,\text{eff}} \approx 5.5\ \mu m^2$, $S_{2,\text{eff}} \approx -20.5\ \mu m^2$ and $S_{3,\text{eff}} \approx 20.5\ \mu m^2$, compared with the areas of $S_1 \approx 5\ \mu m^2$, $S_2 \approx -20\ \mu m^2$ and $S_3 \approx 20\ \mu m^2$ before correction. Therefore, the phases of the single junctions connected to corresponding loops should also be corrected according to the formula $\phi = 2\pi B S_{\text{eff}}/\phi_0$. For the trijunction at $P_1$, the loop area ratios, thus the phase ratios, remains unaffected, i.e., $-1: 1: 2$. (lower left : lower right : central). However, for the trijunction at $P_2$, the ratios become approximately $1: 2.73: 3.73$ (upper left : lower left : lower right).

Secondly, we consider the inductance effect of the superconducting loops. When a supercurrent flows through a loop, it generates additional magnetic flux. Consequently, as the magnitude of the supercurrent varies periodically with the magnetic field, it imposes a periodic modulation on $\phi_{\text{real}}$. For the trijunction at $P_1$, the two associated loops have nominal phases $\phi_2$ and $\phi_3$, with corresponding loops' self-inductance $L_2$ and $L_3$. For the trijunction at $P_2$, the two associated loops have nominal phases $\phi_1$ and $\phi_3$, with loops' self-inductance $L_1$ and $L_3$. The estimated self-inductance of the loops are approximately $L_1 \approx 61$ pH and $L_2 = L_3 \approx 178$ pH. By introducing the parameter $\beta_1 = 2\pi \times I_c L_1/\phi_0 \approx 0.055$, $\beta_2 = \beta_3 = 2\pi \times I_c L_2/\phi_0 \approx 0.17$ (where $I_c \approx 300$ nA), the phases $\phi_{1,\text{real}}$, $\phi_{2,\text{real}}$ and $\phi_{3,\text{real}}$ satisfy the following relations:

$$\phi_{1,\text{real}} = \phi_1 - \beta_1 \sin(\phi_{1,\text{real}}) - \beta_1 \sin(\phi_{1,\text{real}} - \phi_{2,\text{real}}) - \beta_1 \sin(\phi_{1,\text{real}} - \phi_{3,\text{real}})$$
$$\phi_{2,\text{real}} = \phi_2 - \beta_2 \sin(\phi_{2,\text{real}}) - \beta_2 \sin(\phi_{2,\text{real}} - \phi_{3,\text{real}}) - \beta_2 \sin(\phi_{2,\text{real}} - \phi_{3,\text{real}})$$
$$\phi_{3,\text{real}} = \phi_3 - \beta_3 \sin(\phi_{3,\text{real}}) - \beta_3 \sin(\phi_{3,\text{real}} - \phi_{1,\text{real}}) - \beta_3 \sin(\phi_{3,\text{real}} - \phi_{2,\text{real}})$$

where $\phi_1 = 2\pi B S_{1,\text{eff}}/\phi_0$, $\phi_2 = 2\pi B S_{2,\text{eff}}/\phi_0$ and $\phi_3 = 2\pi B S_{3,\text{eff}}/\phi_0$.

The origin of the slight shifts in the main text lies in the difference between the real magnetic fluxes in the superconducting loops – denoted as $\phi_{1,\text{real}}$, $\phi_{2,\text{real}}$ and $\phi_{3,\text{real}}$ – and the nominal magnetic fluxes $\phi_1$, $\phi_2$, $\phi_3$ used in lattice models. To correct the errors from the two previously mentioned sources (flux redistribution and self-inductance modulation), we have modified our calculation by substituting the nominal phases with the real phases. The corrected phases are shown in Figs. S1(a-c), along with the corrected energy-phase relations of the ABS bands for $P_1$ and $P_2$ in Figs. S1(d, g). The new comparisons between $(\alpha, \Gamma)$ and the corrected local density of the in-gap states are shown in Figs. S1(e, f, h, i).

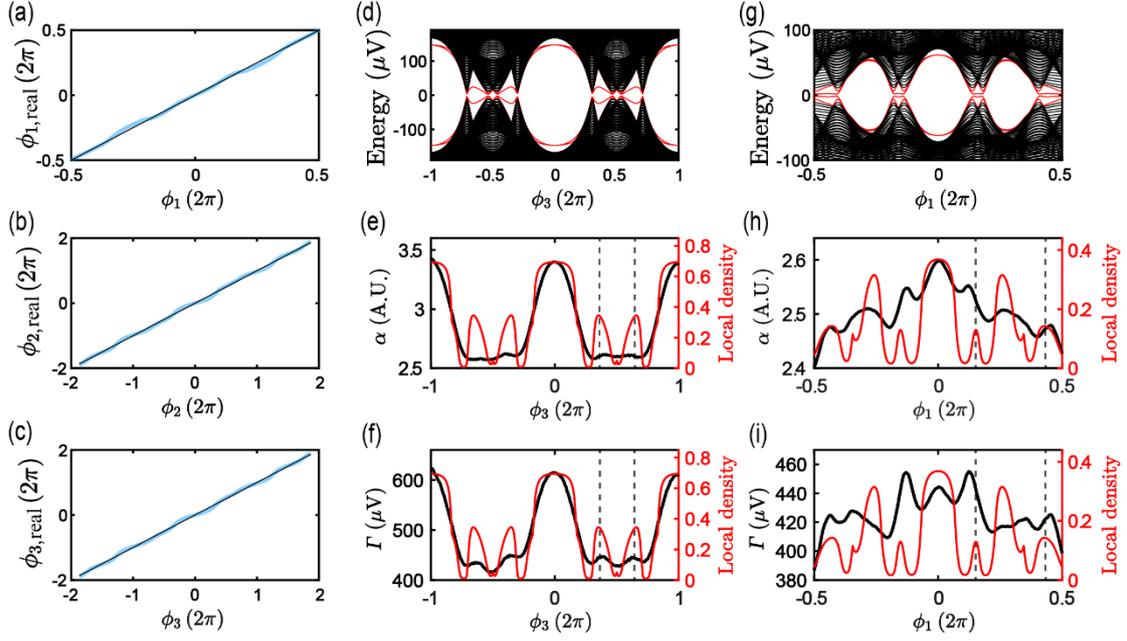

FIG. S1. Flux corrections for the lattice model and the replotted Fig. 4 of the main manuscript after the corrections. (a, b, c) The relations between the real phase $\phi_{1,\text{real}}, \phi_{2,\text{real}}, \phi_{3,\text{real}}$ and the nominal phase $\phi_1, \phi_2, \phi_3$, where the range for $\phi_1 = 1.1\phi$ in the interval of $(-0.5\pi, 0.5\pi)$, $\phi_2 = -3.73 \times \phi_1 = -4.10\phi$ and $\phi_3 = 4.10\phi$, respectively (where $\phi$ is the nominal phase in Fig. 4 of the main manuscript). (d) The calculated ABS bands for the trijunction at $P_1$, obtained by replacing the nominal phase differences $\phi_2$, $\phi_3$ and $\phi_2 - \phi_3$ in the lattice model with the real phases $\phi_{2,\text{real}}, \phi_{3,\text{real}}$ and $\phi_{2,\text{real}} - \phi_{3,\text{real}}$. (e, f) Comparison between the local density of the corrected in-gap states of the trijunction at $P_1$ shown in (d) (red lines) and the $(\alpha, \Gamma)$ coefficients extracted from the experiment data, demonstrating improved alignment in their peak positions after the corrections. (g-i) Similar results for the trijunction located at $P_2$, also demonstrating better alignments between the peak positions.

## 2. The contact conductance measured by the probe electrode at position $P_3$ of the envelope-shaped device

In the envelope-shaped device shown in Fig. 1 of the main manuscript, there was one working probe electrode contacting to one of the single junctions at position $P_3$. In this part of the Supplemental Material we present the contact conductance data measured by that probe electrode and compare the data with the simulation of the lattice model.

Figures S2(a, b) are the 2D maps of measured and simulated contact conductance, respectively. Because the single junction at position $P_3$ is mainly a part of the trijunction at position $P_1$, we therefore use the same ABS bands as that of the trijunction at $P_1$. The two global fitting parameters $\Delta_3 \approx 45\ \mu$eV and $t_3 \approx 217\ \mu$eV, and the $\phi$-dependent fitting parameters $(\alpha, \Gamma)$ shown as the black curves in Figs. S2(e, f), are all obtained in the same way as in the main manuscript.

It can be seen that, the fitted vertical and horizontal line cuts in Figs. S2(c, d), as well as the simulated 2D map of contact conductance in Fig. S2(b), agree well with the experimental data, demonstrating the validity of the simulation including the picture of energy-level-broaden due to coupling.

Plotted in Figs. S2(e, f) are the fitting parameters $(\alpha, \Gamma)$ and the local density of the in-gap states in the single junction at $P_3$. The peak positions of $(\alpha, \Gamma)$ match well with that of the local density. While calculating the local density of the in-gap states, the spatial summation was taken over the four lattice sites (two pairs) at the 25$^{th}$ and the 26$^{th}$ sites away from the trijunction's center. These specific sites roughly represent position $P_3$ in the single junction investigated. We note that in the lattice model each single junction is discretized into 100 unit cells, with a total of 200 lattice sites.

Position $P_3$ may also be interpreted as a part of the upper-left trijunction. However, we approximate it as belonging to the bottom trijunction centered at $P_1$ in our simulations because the spatial weight of in-gap states decays with distance from the trijunction center (Fig. 3(b), main manuscript). Numerical simulations confirm that at distances of approximately 25 unit cells from the bottom trijunction center, the in-gap states remain dominated by this trijunction, with negligible influence from the upper-left trijunction.

The local density peaks of the in-gap states at $\pm 0.125 \times 2\pi$ in Figs. S2(e, f) are noticeably broader than those of $\alpha$ and $\Gamma$ peaks. We attribute this broadening primarily to the Meissner and self-inductance effects within the superconducting loop (discussed in the first part of the Supplemental Material). This is because the single junction at position $P_3$, which is relatively wider than other single junctions and with a larger critical supercurrent, likely distorts more the actual magnetic flux in its associated loop. Accounting for this magnetic flux distortion should narrow the observed peaks in Figs. S2(e, f).

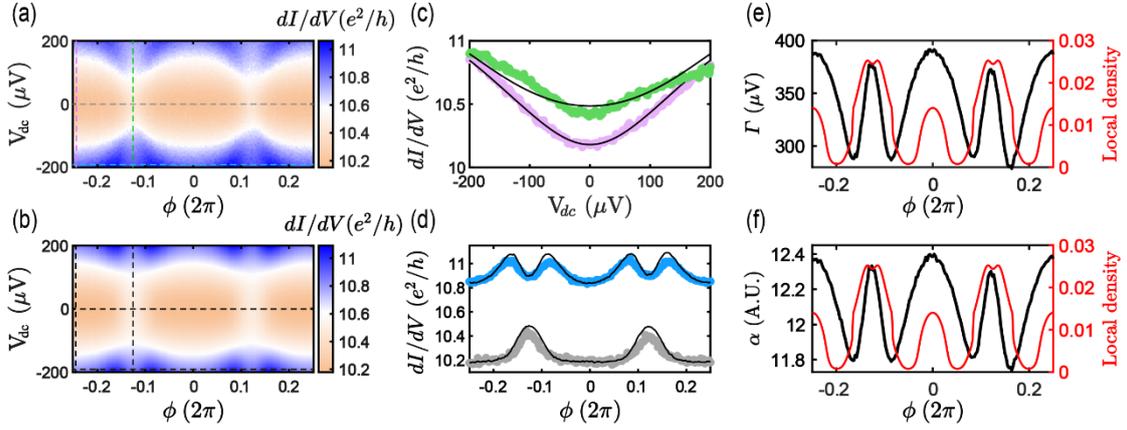

FIG. S2. Experimental data and numerical simulation of the contact conductance measured on the single junction at position $P_3$. (a, b) 2D maps of the measured and simulated contact conductance, respectively. (c) Vertical line cuts of the measured (dots) and simulated contact conductance (black lines) at $\phi = -0.25 \times 2\pi$ (pink) and $\phi = -0.125 \times 2\pi$ (green). (d) Horizontal line cuts of the measured (dots) and simulated contact conductance (black lines) at $V_{dc} = 0\mu V$ (grey) and $V_{dc} = -200\mu V$ (blue). (e, f) Comparison between the local density of the in-gap states of the single junction at $P_3$ and the $(\alpha, \Gamma)$ coefficients extracted from the experiment data, demonstrating a positive correlation in their peak positions.